# How Context Impacts on Media Choice

Stefan Stieglitz
Department of Information Systems
University of Münster
Münster, Germany
Email: stefan.stieglitz@wi-kuk.de

Tobias Brockmann
Department of Information Systems
University of Münster
Münster, Germany
Email: tobias.brockmann@wi-kuk.de

Milad Mirbabaie
Department of Information Systems
University of Münster
Münster, Germany
Email: milad.mirbabaie@wi-kuk.de

**Abstract**

*The relevance of mobile working is steadily increasing. Based on new mobile devices (e.g. smartphones) and their innovative functionalities, an increasing amount of data is being made available ubiquitously. As a result, the growing diffusion of smartphones offers new potential for enterprises. Current mobile devices and related mobile networks have reached a high level of maturity. Thus, the organizational aspects of mobile work have become a focal point of interest for enterprises as well as for academics. This research article addresses the question: How does context influence the choice of communication channels of mobile knowledge workers? An explorative research approach is used to collect and analyse 418 communication incidents, which were initiated by mobile knowledge workers. The results indicate that (1) the context (e.g. travelling on trains) influences the usage of communication channels and (2) smartphones enable the usage of communication channels (e.g. email) in certain contexts.*

## INTRODUCTION

The accelerating diffusion of mobile devices and mobile applications offer new capabilities for supporting business processes in various areas (Houy et al. 2011). In particular, smartphones provide a rich spectrum for mobile workers to increase the efficiency of their business activities. The emergence of so-called "mobile apps" have yielded novel possibilities in the use of smartphones. The smartphone audience in the biggest five states in the European Union (EU5) achieved a significant increase of 44%, up to 104 million subscribers, representing 44% of all mobile users in 2012. In the same period, the US faced an even larger increase of 55% to 98 million smartphone subscribers, representing nearly 42% of all U.S. mobile users (ComScore 2012).

The ubiquitous availability of information through mobile devices has led to an increasing independence of knowledge workers from their stationary workplaces. As a result, the ordinary workplace begins to lose its importance and a growing share of work-related activities takes places outside the office (Venezia and Allee 2007). The support for mobile working by the management of enterprises is expected to increase the satisfaction and productivity of employees while they are working away from the office. CIOs and IT managers are forced to develop a mobile strategy to ensure organizational performance. Hence, they need to know how the mobile workforce communicates while they are on the road (Stieglitz and Brockmann 2012).

These developments raise several new questions about how to handle the support for various devices, how to provide data security, and how to adopt and even manage mobile devices (Ortbach et al. 2014). Mobile workers are challenged by the question of which communication channel works best in relation to specific information needs and tasks. In contrast to media choice decisions at stationary workplaces, there is an additional aspect that seems to strongly influence the appropriateness of a communication channel: "context" (c.f. Table 2).

In this sense smartphones offer mobile knowledge workers a variety of communication channels, for example, email, (video)calls, text messages, and interfaces with software systems. Mobile knowledge workers have to decide which communication channel will probably best satisfy their information needs. Until now, how these decisions are made and what role the business ecosystem (Moore 1996) plays in media choice has hardly been examined. Existing theories in this field do not consider the changing environment of mobile knowledge workers; therefore, they do not provide answers to that question (Grantham and Ware 2009). To make a first step in this direction this paper addresses the following research questions:

1) What is the current position regarding the mobile communication of mobile knowledge workers?

2) What kind of communication channels do mobile knowledge workers use on their smartphones and in what contexts do they use them?

The remainder of the paper proceeds as follows: first we provide an overview of the current literature regarding mobile working (section 2); next, the theoretical background (media theories) is described in section 3; following this, the methodology as well as the results of the empirical study are presented (section 4) and discussed (section 5). The article ends with a summary, which discusses underlying limitations, and provides a perspective on further research (section 6).

## RELATED WORK

In recent years mobile devices have massively invaded the day to day private and business life (Wiredu 2007). The development of mobile devices such as mobile phones and personal digital assistants (PDAs) began in the early 1990s, when mobile phones were designed for communication purposes; however, the technology development proceeded quickly (Kaasinen 2005). Nowadays, research in the field of mobile devices covers technical implementations (Duda et al. 2008), the security of mobile applications (Steele and Tao 2006) and user interface design (Hertzog 2004). Furthermore, some studies have specifically evaluated the use of mobile technologies for business purposes (Gebauer and Shaw 2002; Markova et al. 2007).

According to this literature, the most common reason for organizations to deploy mobile devices is to facilitate information management on the move and thus to allow the mobile workforce to be more productive (Wiredu 2007). According to Saugstrup and Henten (2003) the term "mobile information and communication technology" denotes a hard- and software-specific infrastructure that enables a person to fulfil their tasks while being on the move (Saugstrup and Henten 2003). In this context, the emergence of smartphones has extended and changed the possibilities for location-independent communication and cooperation. As Huth (2011) showed, a stronger conflation of work-related activities and private life can be observed (Huth 2011). Furthermore, based on a study on mobile phone usage data, Wajcman et al. (2008) found that the emergence of mobile information and communication technology has also caused the expansion of working time into private life (Wajcman et al. 2008). According to this, the increased use of mobile phones requires more flexible and better coordination of employees, and it may lead to the increased productivity of the company (Wajcman et al. 2008).

Watson-Manheim and Belanger (2007) discussed the employee's choice of communications media for specific working situations (Watson-Manheim and Belanger 2007) Therefore, they referred to existing media choice theories such as the media richness theory (Daft and Lengel 1986) and the media synchronicity theory (Dennis and Valacich 1999). Watson-Manheim (2007) showed that, for proper media selection, it is not sufficient to consider the comprehensiveness of a certain type of media or the ambiguity of a task. Likewise, it is not enough to merely distinguish between electronic media and face-to-face communication. The results of the study imply that the choice of communication instruments made by individuals depends on various factors, such as the experiences with a certain instrument, the context/location, or the task. Brockmann and Stieglitz (2013) investigated, depending on the tasks, which business processes are requested by mobile workers. They found that mobile knowledge workers are better equipped with the support of communication processes (Brockmann and Stieglitz 2013).

Media choice theories attempt to find answers to why specific media are used to solve specific tasks. Stubblefield et al. (2010) investigated the role of cell phones in the context of new technologies and possibilities. They referred to the social capital theory and the media richness theory in developing a research model to explain the individual's different uses of various features. Complementary to this, Riemer and Filius (2008) stated that a characteristic of the application of mobile systems for business is that both the specific communication and information needs as well as the direct context of the employee (e.g. on travel, in a meeting) have an influence on the success of their usage (Riemer and Filius 2008).

# THEORETICAL BACKGROUND

**Media Choice Theories**

One of the most recognized media choice theories is the media richness theory (MRT). This theory claims that "rich" information is appropriate for equivocal tasks, whereas "less rich" information is suitable for uncertain tasks (Daft et al. 1986; Dennis et al. 2008). Although MRT defines media characteristics well and has been experimentally validated by several studies (Daft et al. 1987; Dennis et al. 1998), many scholars have claimed that recent developments in information and communication technologies merit a more deliberate and redefined form of explanation of the efficacy of various media (Dennis et al. 1999). Additionally, it has to be remarked that some studies have shown that MRT by itself is not sufficient to fully explain media choices (Markus 1994; Mennecke et al. 2000). Markus (1994) argued that social pressures may influence media use much more strongly than richness and in ways that are inconsistent with media richness theory's key tenets. Two approaches to handle this problem are discussed in the literature: (1) refine the MRT to address new communication media (Dennis et. al. 2008) and (2) develop a new theory to consider the requirements of new technologies (Rana et al. 1997).

As a result of this discussion the media synchronicity theory (MST) has been developed to extend the media richness theory for new communication media (Daft et al. 1987; Dennis et al. 1999). In contrast to MRT, the MST states that the richness of a medium is not as crucial as its synchronicity. Dennis et al. (2008) argued that the the fit of media capabilities to the communication needs of a task influences the appropriation and use of media. This in turn influences the communication and the task performance (Dennis et al. 2008). Generally the MST proposes five characteristics of communication capabilities that influence media synchronicity: (1) immediacy of feedback, (2) symbol sets, (3) parallelism, (4) rehearseability, and (5) reprocessability. Furthermore, two types of communication processes determine media usage.

Moreover, media choice is influenced by various factors and it influences communication behaviour (Dennis and Reinicke 2004). Following this line of thought, some media seem to make the interaction simpler for the users. This may depend on the experiences, attitudes, skills, and familiarity of the users, as well as the social norms of the users (DeSanctis and Poole 1994). Media capabilities, tasks, and communication processes influence the media choice, but so do their users' preferences (Dennis et al. 2001). Mobile devices (such as smartphones, tablets, and notebooks) offer highly advanced computing capabilities and connectivity compared to a feature phone. Thus, mobile devices include a broad range of media capabilities in one device, and they support different communication processes. According to the MST, these different media channels have different capabilities (Table 1).

Table 1. Mobile media channels and media synchronicity

|  | Immediacy of feedback | Parallelism | Symbol sets | Rehearse-ability | Reprocess-ability | Synchronicity | Convenience |
|---|---|---|---|---|---|---|---|
| Mobile Calls | High | Medium | Low | Low | Low | High | High |
| Mobile Apps | Medium | High | High | Medium | Medium | Medium | Medium |
| SMS/MMS | Medium | Low | Medium | High | High | Medium | Medium |
| Notifications | Low | Low | Low | High | High | Low | Low |
| Web browsing | Low | High | Medium | Medium | Medium | Low | Low |
| E-mail | Low | Medium | Medium | High | High | Low | Medium |

MRT and MST do not take account of the particular business ecosystem as an "appropriation factor" in making media choices (Ahuja et al. 2007; Riemer and Filius 2008). Appropriation factors are understood to be particular resources, environmental conditions, and situations that affect the usage of media technologies and devices.

**The Context as an Appropriation Factor**

According to Dey et al. (2001), "context is any information that can be used to characterize the situation of an entity. An entity is a person, place, or object that is considered relevant to the interaction between the user and the application, including the user and the applications themselves". Because of this, it is important to consider the given ecosystem, the context, and the mobile communication competence of the user. Schmidt and Forbess (1999) stated that "a context describes a situation and the environment a device or user is in". They differentiated between human factors and physical environments. Human factors describe the identity of a user (e.g. his or her profession), the social environment (e.g. being in a meeting or on a bus), as well as the task (e.g. asking for information or sharing information). Physical environment includes the conditions (e.g. environmental temperature, noise, light), the infrastructure (e.g. web connectivity, power supply), and location (e.g. home-office, customer's office, restaurant).

Mobile devices enable employees to ubiquitously communicate by means of various communication channels. Among other factors, the decision to use a specific communication media may depend on the individual context of the user. The context might be a place (home-office, desk, car), or a certain location, but it may also be specific situations like being at meetings, having spare time, or travelling on a train. On the one hand, the context limits the usage of certain media (e.g. telephone calls in meetings may be perceived as inappropriate); on the other hand, the context assists the choice of a communication media (e.g. sending emails in a train).

To better illustrate the impact of context on media choice we gathered some examples according to Schmidt and Forbess (1999) (see Table 2).

Table 2. Conditions of human factors and physical environment in various contexts

| Context | Human factors | Physical Environment |
|---|---|---|
| Car (when driving) | Attention is needed to manage the traffic, which eliminates the possibility of writing or reading. | Connectivity might be temporarily insufficient. Often no power supply is provided. Hands-free device is necessary. |
| Train | There might be some social pressure to avoid phone calls while sitting in a train. | Connectivity might be temporarily insufficient. Space may be limited. Often, no power supply is provided. Noises or light conditions may disturb working activities. |
| Plane | - | No connectivity. Space may be limited. Often no power supply is provided. |
| Walking | Awareness is needed to obstacles and traffic. While moving it is difficult to write. However, it is still be possible to make phone calls or to read. | Good connectivity. Noises may disturb communication activities (e.g. phone calls). No power supply is provided. Messaging/typing is difficult. |
| Meeting | Attention is needed to follow and participate in the discussion in the meeting. There might be some social pressure to avoid phone calls while in a meeting. | Often good connectivity and power supply. |
| Waiting | Communication processes may be disturbed abruptly. Working while waiting is an efficient use of time. | Depends on location. Often no power supply |
| Hotel | Working conditions are similar to those in a stationary working space. | Good connectivity and power supply. |
| Home office | Working conditions are similar to those in a stationary working space. | Good connectivity and power supply. |

As we have shown, the existing media choice theories do not sufficiently explain the communication behaviour of mobile workers. This type of employee obviously has to consider his or her location to choose the best communication channel. Therefore, we suggest extending the MST by adding the attribute *context* as a new "appropriation factor" (c.f. Figure 1).

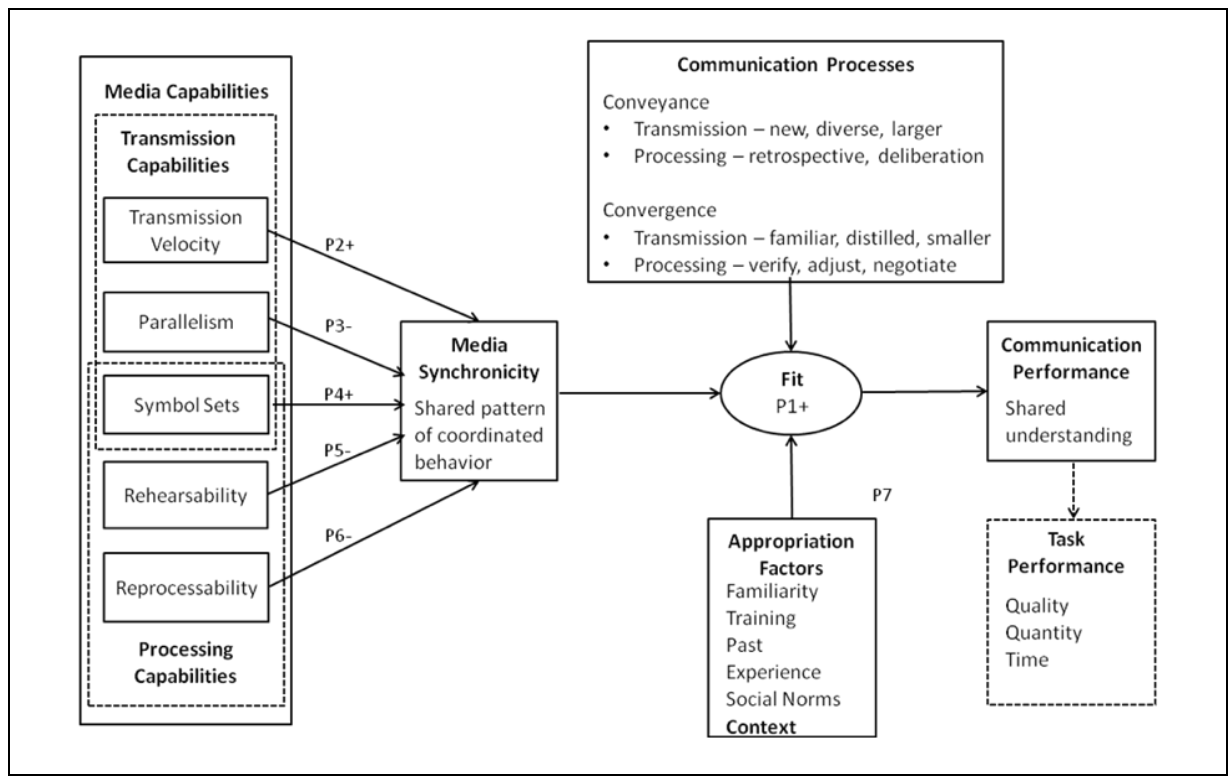

Figure 1: Media synchronicity theory according to Dennis et al. (2008)

# EMPIRICAL STUDY

**Methodology**

We set up our research design, in order to analyse the mobile communication and information behaviour of employees within companies. Our aim was to learn more about the preferences of mobile knowledge workers for different media channels in certain contexts. Since there is only very limited research in this field, we decided to conduct an explorative empirical study. One major goal of this study was to obtain information about the coherence of the communication channel used and the situation (context) of mobile knowledge workers.

We adopted a research methodology that enabled the participants to immediately record perceptions (such as context) and activities. The participants were required to continuously capture their communication and information-sharing behaviour over a period of two to three days. Therefore, each participant obtained a pre-prepared paper-based notebook for the evaluation period in order to note each mobile communication incident. For each incident, a sheet had to be completed. The sheets were designed to be as intuitive as possible in order to raise the response rate. We first developed a test form and then improved it according to the requirements of the target audience through an iterative process by conducting five semi-structured interviews as well as 10 beta-tests with consultants (i.e. for testing if all frequently found contexts were mentioned on the sheets).

Between March 2013 and July 2013 we analysed the communication behaviour and media usage of 42 randomly selected business consultants from different companies, who mainly worked outside their firms' offices (>75% of their working time). While on the road they were equipped with smartphones (100%), notebooks (83%), and, in some cases, with tablet PCs (22%).

**Results**

Due to the small number of observed consultants and the obvious gender bias the research results should be interpreted and generalized with great caution. Table 3 shows the demographics of our study.

Table 3. Descriptive statistics

| | |
|---|---|
| Number of participating consultants | 42 |
| Number of recorded communication acts | 418 |
| Gender | 89% male; 11% female |
| Average age | 33 years |
| Technical equipment | Smartphone: 100 %<br>Laptop: 83%<br>Tablet: 22% |
| Working time spent outside the office (average) | >75% |
| Completed communication sheets on average | 10 per consultant |

In the following, we provide general information about the communication behaviour of mobile knowledge worker. This information should help the reader get an impression of how the participants communicated. Overall, all this information answers the first research question: "What is the current position regarding the mobile communication of mobile knowledge workers?"

- The most frequently used mobile devices were smartphones (59%), followed by laptops (34%). The evaluation shows that tablets (2%) were practically never used, even when available.

- The analysis of the attribute "communication channels" indicates that emails (35%) and phone calls (30%) were the most frequently used channels. In 14% of all cases the participants were connected to intranet systems.

- The use of mobile applications (7%), access to websites (5%), and the use of text messages (5%) played a minor role.

- The cancellation of a process (waiver of communication) was reported in 7% of all cases.

- The evaluation shows that 43% of the participating consultants requested some kind of information in their communication acts, whereas approximately one third (36%) wanted to share information with others.

- The majority of all the communication incidents examined can be described as being of an organizational (27%) or project-specific (25%) nature. Scheduling (18%) and the exchange of personal information (14%) played a minor role. The work on status reports (7%), customer information (5%), and billing (4%) was negligible.

- The priority of the communication incidents was classified as moderate, medium, or high. In 16% of all cases the priority of a communication incident was denoted as low, in 57% as medium, and in 28% as high.

As mobile workers travel to and from work in different places, we included situational context as an influencing factor in our analysis. Surprisingly, the analysis reveals that the most frequent context for initiating communication incidents using mobile IT solutions was "meeting" (29%), while 19% took place in trains, and 12% in cars. More rarely mentioned were "by foot" (9%) and in "waiting" situations (7%). The term "others" was used in 18% of the cases. Additional notes by the participants indicated that "others" mostly referred to so-called "third places" such as cafeterias, hotels, restaurants, coffee kitchens, and office corridors.

To gain a deeper insight into the communication activities of the participants with respect to the information and communication incidents, and to answer research question 2, a contingency analysis (Table 4) was conducted to determine the statistical coherence between each attribute. Pearson's chi-squared was used to assess the extent of coherence.

Table 4. Contingency analysis of the influence factors in mobile communication

|  | Information flow | Priority | Device | Context | Channel | Content | Partner |
|---|---|---|---|---|---|---|---|
| Information flow | 1.00 | 0.48 | 0.24 | 0.24 | 0.18 | 0.20 | 0.28 |
| Priority |  | 1.00 | 0.10 | 0.26 | 0.27 | 0.30 | 0.30 |
| Device |  |  | 1.00 | 0.54 | 0.60 | 0.28 | 0.13 |
| Context |  |  |  | 1.00 | 0.53 | 0.33 | 0.50 |
| Channel |  |  |  |  | 1.00 | 0.43 | 0.44 |
| Content |  |  |  |  |  | 1.00 | 0.48 |
| Partner |  |  |  |  |  |  | 1.00 |

Next, we will describe the four relations with the highest chi-square coefficient (>0.5): device–channel, channel–context, device–context and partner–context (shaded grey in Table 4).

Table 4 shows that the choice of a device and the communication channel is highly correlated, with a chi-square coefficient of 0.6. There is a quite simple reason for this correlation: smartphones were used exclusively for all phone calls and text messages. Enterprise systems, which were accessed in 7.9% of cases for communication, were exclusively accessed via notebooks. However, laptops were used more intensively than smartphones in meetings.

A high correlation can also be observed between "communication channel" and "context", with a chi-square coefficient of 0.53. Context describes, for example, situations in trains and cars, or meetings at partner firms. The high coefficient can easily be explained. For example, consultants exclusively made calls using their smartphones when travelling in cars because they did not have any other means of communicating while driving. In total, 29% of all communication incidents happened during meetings. On 45% of these occasions consultants used email, and in 32% they used their smartphones for phone calls (during meetings). A reason for this could be that information in communication incidents that occurs during meetings results from the meeting itself and therefore has a high situation-related priority, which necessitates communication.

The choice of a device and context is highly associated, with a chi-square coefficient of 0.54. Overall, two factors might influence the relation between context and device. In the contexts of "car", "train", "walking", "waiting", "meetings (with colleagues)", and "others", the smartphone was the device predominantly used. However, in "meetings (with customer)" and "home-office", the notebook was the device that was mainly used. The relatively strict dependence explains the strong coherence. Therefore, it is useful to distinguish between smartphone- and laptop-suited contexts. Furthermore, "subject-specific" and "scheduling activities" are primarily accomplished by smartphones during the smartphone-suited contexts.

There is a relatively distinct coherence between "communication partner" and "context" (0.50). It is not surprising that "communication with colleagues" in the context of "home-office" and "meetings" is higher. More interesting, is that communication with supervisors mainly happens in cars as well as in the context of the category "others", such as in coffee bars, on elevators, and so on. Communication with "customer" is highest while "walking".

## DISCUSSION

In our second research question we asked about the communication channels that mobile knowledge workers used on their smartphones and the contexts in which smartphones are used. Our analysis proves that there is a strong relationship (0.53) between the direct environmental situation (context) and the mobile knowledge workers' choice of a specific communication channel. This finding enables us to state that the choice of communication channel depends on the context. To provide a few more insights from our study, we found that phone calls mostly occurred in cars (34%), followed by meetings (18%). Additionally, based on the theoretical discussion, we supposed that the social environment might prevent participants making phone calls in specific contexts (e.g. on a train or while at a meeting). However, this seems not to be the case. There are two possible explanations of this phenomenon: first, half of the meetings took place internally, which means that only work colleagues participated in the meeting. In this situation the social restrictions that discourage phone calls might be lower than in, for example, a meeting with customers. Second, the diffusion of mobile technologies and their ubiquitous usage might have increased the acceptance of making phone calls anywhere and at any time, which might also apply in other contexts such as on trains or in restaurants.

A great challenge for communication via notebooks is the limited access to broadband networks and the amount of time required to set up the device. In this case, it can be claimed that smartphones have clear advantages. For instance, they offer access to emails and, furthermore, they are normally always online. Our study confirms this assumption: nearly 42% of the email communication via smartphone is done in the contexts of train and walking. This addresses the issue raised in our second research question.

Enterprise systems, as well as web resources and intranet systems, were rarely used by our mobile knowledge workers. Only a few participants used their smartphones for these purposes, and these were mostly within the contexts of "waiting", "other", and "hotel". It can be assumed that until now interfaces with enterprise systems and intranet sites are not optimized for mobile browsers, which limits their usability and the communication possibilities of such systems. Furthermore, the usability of complex websites is rather low. Mobile apps (33%) as well as text messages (35%) are mainly used in trains.

These results demonstrate that the media synchronicity theory (Dennis et al. 2008) should be extended by the addition of the appropriation factor, "context". In essence, our study demonstrates that smartphones are used in various contexts to a greater or lesser extent, despite the specific conditions of the human factors and the physical environment. Therefore, it seems to be true that smartphones are used as all-purpose devices, while feature phones and laptops usually have stronger limitations on usage, based on the context. However, the usage of communication channels like apps, enterprise-systems, and websites is at a low level. One explanation might be that there is low support from the IT departments for the strategic integration of smartphones into enterprise systems. However, this might change significantly within the next few years, making apps a promising approach to support mobile workers' communication and collaboration.

## CONCLUSION

Mobile communication and information acquisition behaviour in the business context is about to change – mainly due to the increasing dissemination and greater functionality of smartphones. The emergence of mobile devices and a mature mobile infrastructure has increased the relevance of mobile work. A broad strand of theories about media usage such as media choice theories exist, including the media richness theory or the media synchronicity theory. Both theories lack situational factors as explaining variables.

In this context, our study is a first explorative attempt focused on gaining a better understanding how mobile knowledge workers use their smartphones in different contexts. After outlining the theoretical fundamentals and the current state of research, we presented the results of an empirical study, which is based on 418 sheets detailing communication incidents completed by 42 mobile knowledge workers. Our results reveal that smartphones are used in various contexts and that different communication channels (e.g. phone calls and email) are accessed via smartphones. We identified that attributes of the context might affect the usage of smartphones. Moreover, we were also able to detect that the potential of smartphones has not yet been fully exploited (e.g. provision of interfaces to enterprise systems).

In contrast to the minor grab sample, the contingency analysis follows an explorative approach. It has been shown that this research design is a suitable one for shedding light on the research question. A key finding of this research is that smartphones enable communication channels to be used in different contexts.

Our article clearly contributes to the academic discussion of human behaviour in IT adoption and use by providing rich insights into decisions about media choice made by mobile knowledge workers. We have provided a theoretical framework by discussing the relevance of the attribute "context" as an appropriation factor of media choice (especially) on mobile devices. Therefore, we suggest extending the MST in this direction. As we have argued in our literature review, there is a significant lack of empirical data in this field. In future research we propose to carry out comparable studies to validate the results and measure the impact of different cultural environments.

## COPYRIGHT